# Exploring AI-powered Digital Innovations from A Transnational Governance Perspective: Implications for Market Acceptance and Digital Accountability

**Claire Li**
*Royal Holloway, University of London*
**David Peter Wallis Freeborn**
*Northeastern University London*

*Completed Research*

**Abstract**

*This study explores the application of the Technology Acceptance Model (TAM) to AI-powered digital innovations within a transnational governance framework. By integrating Latourian actor-network theory (ANT), this study examines how institutional motivations, regulatory compliance, and ethical and cultural acceptance drive organisations to develop and adopt AI innovations, enhancing their market acceptance and transnational accountability. We extend the TAM framework by incorporating regulatory, ethical, and socio-technical considerations as key social pressures shaping AI adoption. Recognizing that AI is embedded within complex actor-networks, we argue that accountability is co-constructed among organisations, regulators, and societal actors rather than being confined to individual developers or adopters. To address these challenges, we propose two key solutions: (1) internal resource reconfiguration, where organisations restructure their governance and compliance mechanisms to align with global standards; and (2) reshaping organisational boundaries through actor-network management, fostering engagement with external stakeholders, regulatory bodies, and transnational governance institutions. These approaches allow organisations to enhance AI accountability, foster ethical and regulatory alignment, and improve market acceptance on a global scale.*

## 1.0 Introduction

This study examines how the Technology Acceptance Model (TAM) can be adapted to evaluate AI innovations in a transnational context. It provides insights into the reasons for organizations to develop and adopt AI technologies, and the ways in which they can navigate regulatory, ethical, and cultural landscapes. Additionally, it explores how organisations can enhance global market acceptance and strengthen accountability for digital innovations.

Technology is one piece of the puzzle that must be solved for companies to remain competitive in a digital world (Rivard 2004). Organisations also require adequate strategies (Bharadwaj et al 2013; Matt et al 2015) as well as suitable internal structures (Selander and Jarvenpaa 2016), processes (Carlo et al 2012), and culture (Karimi and Walter 2015) to yield the capability to generate new paths and innovations for value creation (Svahn et al 2017). Existing studies offer valuable insights into value creation (Vial, 2021), yet the organisational motivations behind innovation development and strategies for improving the process remain underexplored. Moreover, there is a lack of understanding regarding how these influences persist after digital innovations have been adopted in other nations. Therefore, our study addresses two research questions from a transnational governance perspective: "*Why do organisations develop and adopt AI powered digital innovations?*" and "*How do organisations manage the process of improving the market acceptance of their AI-powered digital innovations?*".

To address these gaps, our study employs TAM as an underlying analytical framework. TAM, initially developed by Davis, Bagozzi and Warshaw (1989), provides a theoretical framework for analysing the factors that influence users' adoption of new technologies. However, the transnational nature of AI innovations introduces complex governance challenges, arising from varying regulatory frameworks and cultural norms. Cultural differences can shape different individuals' perspectives and priorities, leading to significantly varying values and approaches across different regions (Toon, 2024). AI developers may unwittingly bring their cultural perspectives and cognitive biases into the process of AI development, for example by using unrepresentative training data or an imperfect algorithmic structure (Fazelpour and Danks, 2021, Athota et al., 2023). Lacking complete information can cause bias by excluding certain groups or sections of the population. The rise of *transnational governance* (Djelic & Quack, 2007; Djelic & Sahlin-Andersson, 2006) has heightened the need for a more nuanced understanding of how AI-driven innovations are accepted across borders, taking into account the diversity of cultural, ethical, and regulatory environments.

This theoretical lens helps us explain how certain factors of digital transformation become critical in market acceptance in transnational governance. However, it neither tells us about the motivations that drive organisations towards digital transformation, nor how organisations can respond to these factors, leading to an increase in market acceptance and digital accountability. Social pressures are especially important in driving organisations toward digital transformation (Gegenhuber et al 2022; Saarikko, Westergren and Blomquist 2020; Zhu et al 2006). When organisations face social pressure, they are more likely to prioritise digitalisation to maintain their legitimacy, reputation, and market position (Lee, Pak and Roh 2024). Additionally, responding to societal expectations can improve corporate social performance, making organisations more responsible and adaptive to new digital trends. In times of social or economic crises, organisations often accelerate digital initiatives to meet new demands, demonstrating that social pressure is a significant catalyst for transformation (Khurana, Dutta and Ghura 2022). We use the concept of social pressure to understand the institutional motivations, particularly regulatory, ethical and cultural pressures, for organisations to develop and adopt AI-powered digital innovations.

Based on these insights, our study proposes two solutions to enhance the transnational accountability and market acceptance of AI-powered digital innovations: reconfiguring internal governance and reshaping organisational boundaries through actor-network management. Drawing on Latour's (1987, 2005) actor-network theory (ANT), which views technological systems as shaped by the interactions of both human and non-human actors, we argue that accountability is co-constructed through dynamic relationships between AI developers, regulatory bodies, users, and governance institutions.

ANT helps to illustrate that AI-powered innovations do not operate in isolation but are embedded within broader socio-technical networks that influence their development, deployment, and societal impact. In this light, organisations must actively engage with these networks to ensure ethical and regulatory alignment. We contend that AI or digital innovations themselves are not inherently biased or culturally insensitive; rather, bias arises from the ways in which organisations develop, train, and direct AI systems. Therefore, from a transnational governance perspective, organisations must assume transnational accountability for digital innovations, as they control the AI development processes, training platforms, and systemic design choices that directly affect market acceptance and ethical considerations. By integrating ANT with transnational governance principles, we highlight the importance of both internal organisational restructuring and external boundary management in fostering responsible, compliant, and globally accepted AI innovations.

Our study aims to make the following contributions to the literature on digital transformation. First, we investigate why and how organisations develop and adopt AI-powered digital innovations in transnational governance. Prior research has focused on how organisations develop digital technologies by building up information architecture (Tan, Abdaless and Liu 2018; Tan, Liu and White 2013). Related studies have explored how digital innovations can become affordable and acceptable by the market via organisational semiotics (Pan et al 2018; Hafezieh and Eshraghian 2022; Hafezieh and Pollock 2023; Nambisan et al 2017). We differ from this stream of literature by exploring how organisations can adapt their AI-powered innovations in a transnational context. In this vein, we further demonstrate how organisations show differences in AI-powered digital transformation when facing different regulatory and ethical pressures. We contend that regulatory, ethical and cultural acceptance pressures are important factors in technology acceptance in transnational governance.

Second, we contribute to the existing literature on the motivations driving organisations toward digital transformation. Previous research has highlighted several key factors that influence the adoption of digital innovations, including perceived ease of use, perceived usefulness, trust, security risks, costs, privacy concerns, cultural context, and social influence (Pan et al., 2018; Kim, Mirusmonov, & Lee, 2010; Shin, 2010; Koenig-Lewis et al., 2015; Lu et al., 2011; Arvidsson, 2014; Slade, Williams, & Dwivedi, 2013; Mallat et al., 2009). We add another layer to the literature by explaining that regulatory, ethical and cultural acceptance pressures are important factors driving digital transformation, drawing ethical AI principles and cultural acceptance discussions.

Third, we respond to Vial's (2021) call to explore the ways in which organisations improve their market acceptance in digital transformation. Prior research has used search as an approach to inform organisations to renew their innovative products, innovation processes and redefine their value propositions (Hafezieh and Eshraghian 2022; Hafezieh and Pollock 2023). We extend this discussion by arguing that digital innovation developers and adopters must assume transnational accountability for their innovations, as AI technologies operate within complex socio-technical networks that transcend national borders. Drawing on Latourian actor-network theory (ANT), we highlight how AI-powered innovations are shaped not only by technological advancements but also by interactions between regulatory bodies, industry stakeholders, and end-users. Accountability, therefore, is not static but co-constructed within these evolving actor-networks. We accordingly propose two solutions for organisations to enhance market acceptance and accountability in AI-powered digital transformation – internal resource reconfiguration and reshaping organisational boundaries through actor-network management.

## 2.0 Motivations of Developing and Applying AI-powered Digital Innovations

### 2.1 Profit-driven Motivation

Organisations are increasingly adopting AI-powered digital innovations to enhance their profitability (Fountaine, McCarthy and Saleh 2019). These technologies allow businesses to automate routine tasks, reduce operational costs, and increase efficiency. AI systems can process vast amounts of data, enabling more accurate forecasting and decision-making, which directly contributes to improved financial performance (Olan et al 2022). Moreover, AI-driven innovations offer personalized services to customers, enhancing customer satisfaction and driving revenue growth (Usman et al 2024). The competitive advantage that AI provides has become a significant motivation for organisations to continuously invest in and develop such technologies.

### 2.2 Market Acceptance and Technology Acceptance Model (TAM)

Market acceptance refers to the process by which consumers or businesses adopt and use a product or technology (Gao et al 2013). It is crucial because it determines the commercial success of a product, influencing profitability and long-term sustainability. Without market acceptance, even the most innovative products may fail due to low adoption rates.

The Technology Acceptance Model (TAM) provides a framework for understanding how AI-powered innovations gain market acceptance (Davis 1989; Silva 2015). TAM is an information systems theory developed to explain how users come to accept and use technology. It is based on two key variables. The first variable is *Perceived Usefulness*, referring to the degree to which an individual believes that using a particular technology or system will enhance their job performance or productivity. If a user perceives that a system or technology is useful and can improve their tasks or roles, they are more likely to accept and use it. This concept highlights the importance of demonstrating clear benefits to users in order to encourage technology adoption. AI-powered innovations are adopted more readily if users perceive that the technology improves their performance or adds value. In fields like accounting, AI tools that enhance efficiency and accuracy can lead to higher acceptance rates.

The second variable is *Perceived Ease of Use*, refers to the degree to which a person believes that using a particular technology or system will be free of effort. If a user perceives that a system is simple and straightforward to use, they are more likely to accept and use it. Perceived ease of use is essential because it reduces the cognitive load and learning curve associated with new technologies, increasing the likelihood of widespread adoption. The simpler and more intuitive AI technologies are to operate; the more likely users are to adopt them. Advanced user interfaces and seamless integration into existing workflows reduce friction in adoption, leading to broader market acceptance.

In the case of AI-powered digital innovations, organisations perceive these technologies as highly useful due to their ability to automate complex processes and deliver real-time insights. At the same time, advancements in user interfaces and AI integration make these innovations easier to use, thereby fostering broader acceptance within organisations.

**2.3 Extended TAM and Transnational Governance**

Surprisingly, despite the increasing importance of AI-powered digital innovations, there has been relatively little exploration of its institutionalisation impacts, particularly in the context of transnational economic and market governance (Arnold 2009a; 2009b; Mehrpouya and Salles-Djelic 2019; Friedrich, Kunkel and Thiemann 2024). Our study extends TAM by exploring the institutionalisation impacts of AI-powered digital innovations from the transnational governance perspective.

The *transnational governance perspective* refers to the collaborative efforts of countries, organisations, and stakeholders to regulate, manage, and oversee emerging technologies or global issues beyond national borders (Roger and Dauvergne 2016). This perspective recognizes that challenges such as AI, climate change, and cyber security require international coordination and governance mechanisms due to their global impact.

In the context of AI, transnational governance involves creating frameworks that standardize rules, ensure ethical usage, and mitigate risks across different jurisdictions. It relies on cooperation between states, international organisations, and private sector actors to establish regulations that reflect shared values, such as accountability, transparency, and fairness. For instance, the European Union's regulatory model for AI is one of the leading examples of successful AI governance at the transnational level. Transnational governance is critical because AI's global nature means that decisions in one region can have ripple effects worldwide. This necessitates diplomatic engagement, international law, and shared policy frameworks to manage both the risks and opportunities of AI.

An extended version of TAM can be applied to consider factors relevant to transnational governance when examining AI-powered innovations. While *Perceived Usefulness* and *Perceived Ease of Use* remain critical, institutional variables such as *regulation compliance* and *cultural and ethical acceptance* should be considered. Organisations operating across borders must ensure that their AI systems comply with various international regulations, such as data privacy laws (e.g., GDPR in Europe). The ease with which AI innovations can be adapted to meet these governance requirements can significantly impact their acceptance. Furthermore, organisations must consider the cultural context in which these technologies are deployed, ensuring they are compatible with local norms and ethical principles.

**2.4 Social pressures as motivations**

Our study uses the concept of *social pressures* to explore the motivations for organisations to develop and adopt AI-powered digital innovations. *Social pressures* refer to the external influences exerted by society, stakeholders, or peers on organisations or individuals to conform to certain norms, regulations, or ethical standards (Bursztyn and Jensen 2017). These pressures can arise from regulatory bodies, customers, cultural expectations, and societal values, prompting organisations to align their strategies and practices with accepted norms to maintain legitimacy, reputation, and competitiveness.

**2.4.1 AI Regulations and Digital Innovation**

Organisations face growing social pressure to comply with AI regulations, which are being developed across different jurisdictions (OECD, 2021). Regulatory frameworks focus on issues such as data privacy, transparency, and accountability in AI development. Failure to adhere to these laws can result in penalties, reputational damage, and loss of consumer trust. Therefore, organisations are motivated to develop and adopt AI innovations that comply with these regulations to maintain market access and ensure sustainable growth.

AI regulations play a crucial role in shaping the trajectory of digital innovations by setting the boundaries within which AI systems operate. Effective regulation is essential to ensure that AI advancements are aligned with societal values and ethical standards, thereby fostering trust and widespread adoption.

AI regulations can significantly influence the pace and direction of digital transformation. Rules-based regulations, such as the European Commission AI Act, passed on March 13, 2024 (EU, 2024), provide certainty and clarity, offering uniform applications across the EU. This approach ensures that AI systems meet specific predefined standards, thus safeguarding public interests and maintaining high levels of trust. However, the rigidity of rules-based regulations can stifle innovation and hinder technological advancements, as they may not be adaptable to the rapid pace of AI development.

On the other hand, principles-based regulations such as those proposed in the UK's White Paper on AI regulation (UK Government, 2023) adopt a more flexible and adaptive approach. This framework focuses on desired outcomes rather than specific rules, encouraging innovation while addressing key principles such as safety, security, robustness, transparency, fairness, accountability, and governance. This flexibility encourages rapid adaptation to new technological developments and diverse applications of AI. However, the inherent ambiguity and subjectivity in principles-based regulations can lead to inconsistent interpretations and applications, posing challenges for legal settlements, and potentially allowing unethical behaviour.

Regulations/rules compliance is a social norm of crime avoidance and can form social pressures for organisations to comply with AI regulations/rules. Incompliance with AI regulations can result in possible crimes and negative social influences for organisations, reducing their technology acceptance level. For this reason, we contend that AI regulation compliance is an important factor to motivate organisations to develop and adopt AI-powered innovations because social pressures push them to comply with social norms and regulations.

By grounding regulations in a robust theoretical understanding of risk, it is possible to create more effective, adaptable, and ethically sound AI governance structures that support sustainable digital innovation.

**2.4.2 Ethical and cultural acceptance of AI-powered innovations**

Ethical and cultural acceptance is another critical social pressure driving organisations to adopt AI responsibly (Lobschat et al 2021). Societies increasingly demand ethical AI usage, with concerns around fairness, bias, and job displacement. Companies that address these ethical concerns may be seen as more socially responsible, and thus more likely to build public trust and customer loyalty. This may motivate organisations to develop AI systems that align with ethical and cultural standards, ensuring acceptance and integration into broader society.

There is widespread agreement that AI regulatory frameworks, whether rules-based or principles-based, should incorporate ethical principles into their design (Ashok et al 2022). These ethical principles are needed to provide a normative foundation for AI governance and to define what constitutes responsible development and use of AI systems. Yet, there are substantial disagreements about which precise ethical principles should be incorporated into regulations governing AI. These disagreements stem from underlying cultural differences, ethical disagreements, as well as uncertainties about how to respond to a rapidly changing technological landscape. Key areas of contention include balancing risks and opportunities, privacy versus security, and transparency against system efficacy and integrity.

Nonetheless, Jobin et al. (2019) identify an emerging consensus around five core ethical principles in guidance documents for ethical AI. We outline these principles below.

The first principle is *Transparency,* meaning that AI systems should be explainable, interpretable, and open to human scrutiny (Zednik, 2021). However, the level and nature of transparency may vary depending on the stakeholder and the specific application. For example, an AI spam-email detection system might offer high transparency to regulators ensuring proper data use, but more limited transparency to individuals to prevent exploitation of system vulnerabilities.

The second principle is *Justice, Fairness and Equity,* meaning that AI systems should not discriminate or create unfair outcomes for different groups (Johnson, 2020; Fazelpour and Danks, 2021). In particular, AI should not perpetuate or exacerbate existing societal biases and inequalities. For instance, a responsible AI recruiting tool should avoid discrimination, whether direct or incidental, based on protected characteristics like gender or ethnicity.

The third principle is *Non-maleficence,* meaning that AI should not cause harm to humans, either deliberately or inadvertently (Floridi and Cowls, 2019). This demands rigorous risk assessment and robust safety measures. For example, an autonomous vehicle should be designed with protection against deliberate or dangerous misuse and multiple fail safes to prevent accidents.

The fourth principle is *Responsibility,* meaning that AI systems should be accountable to humans, and responsible to human oversight (Matthias, 2004). This may involve clear liability frameworks for AI developers and specific human oversight requirements. For instance, AI-driven financial trading systems should have a well-defined human accountability chain for errors, even when decisions are made autonomously.

The fifth principle is *Privacy,* meaning that AI systems must be responsible in their use of personal data and information (Nissenbaum, 2010). This might involve data protection measures and respect for privacy rights. For example, a responsible AI financial assessment tool should implement strict safeguards to protect users' personal data, potentially including data anonymisation or deletion protocols.

These ethical principles provide a framework for identifying the ethical problems that arise in the context of digital innovations driven by AI. However, these principles can sometimes conflict with each other (Blanchard et al., 2024; Sanderson et al., 2023). For example, transparency and privacy may conflict in credit rating algorithms. While individuals and regulators may request explainability in the outcomes, strict privacy laws may demand data minimisation, limiting what can be disclosed. Similarly, fairness and non-maleficence can create conflicts in AI hiring tools. Algorithms designed to ensure demographic fairness may adjust selection criteria, but this can reduce accuracy, leading to less optimal hiring decisions. These inherent tensions will often necessitate some tradeoffs between the ethical principles.

AI systems are not merely technical tools; they are embedded in cultural and ethical contexts that vary across different societies, and AI-driven digital innovations are becoming more integrated into daily life. As such, organisations need to navigate the challenge of aligning their AI innovations with diverse cultural expectations, whilst also addressing ethical concerns. Under this social expectation, a form of *accountability* emerges for organisations that create or deploy AI. Rather than being solely driven by regulatory compliance, companies are increasingly compelled to take responsibility for how their AI-driven platforms influence culture, identity, and representation. Public scrutiny, ethical debates, and the demand for responsible AI have pushed organisations to be more transparent about their design choices, data sources, and decision-making processes. This accountability is not just about preventing harm; it is also about gaining legitimacy and trust. When organisations demonstrate ethical responsibility in AI development, they enhance their reputation and increase the likelihood of widespread acceptance of their technologies.

What makes this accountability even more significant is that it transcends national borders, creating *transnational accountability* for digital innovations. AI is a global phenomenon—its impact is not confined to the country where it is developed but extends across multiple regions and cultures. As a result, we contend that the responsibility for ensuring ethical AI use does not belong to any single government or organisation. Instead, multinational organisations, international regulatory bodies, and cross-border advocacy groups all play a role in shaping AI governance. Transnational accountability arises when organisations must answer not only to their home governments but also to global stakeholders, including consumers, civil society organisations, and policymakers from different countries. This interconnected accountability structure drives the push for shared standards, ethical frameworks, and collaborative governance efforts to ensure that AI is developed and used in ways that respect cultural diversity and human rights worldwide.

We contend that AI regulations, ethical and cultural acceptance form the main social pressures for organisations to develop and adopt AI-powered digital innovations. From a transnational governance perspective, we propose that this social pressure not only motivates organisations in technology advancement and market acceptance but also creates mutual and transnational accountability for digital innovations.

Based on these discussions, we propose the following framework of motivations for organisations to develop and adopt AI-powered digital innovations (Figure 1).

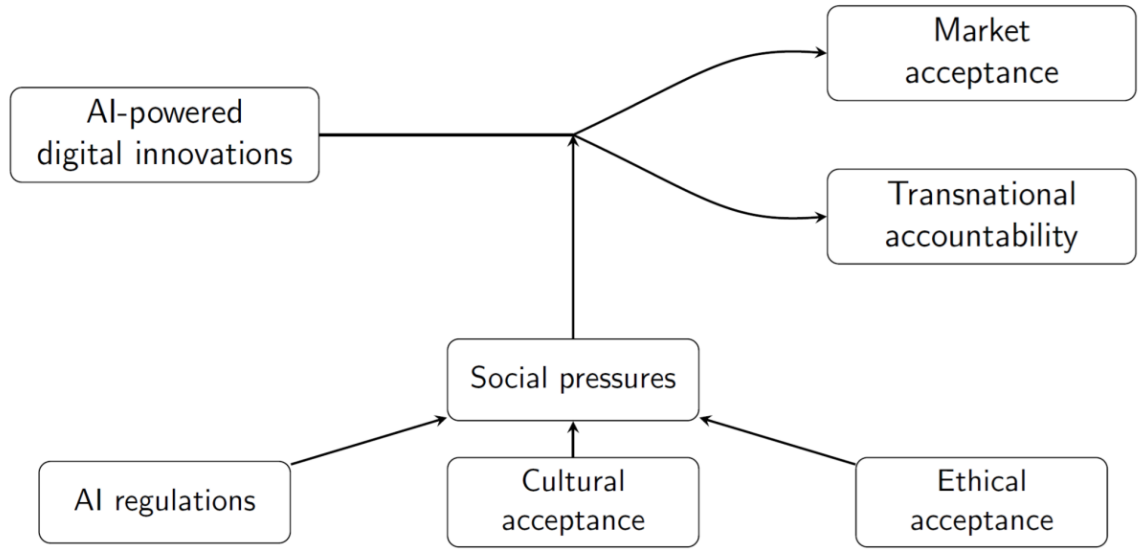

**Figure 1. Motivations for organisations to develop and adopt AI-powered digital innovations.**

## 3.0 Proposed solutions: Enhancing Transnational Accountability for AI-powered Digital Innovations

### 3.1 Theoretical Foundations: Latourian Perspective, TAM, and Transnational Governance

We propose solutions based on the previously discussed theoretical foundations, the Technology Acceptance Model (TAM), transnational governance, and Latourian actor-network theory (ANT). Each provides a complementary lens for analysing how organisations navigate AI accountability and market acceptance across global contexts.

From the perspective of ANT, accountability is not a fixed or top-down process but is instead co-constructed through interactions between human and non-human actors (Latour, 2005). In AI governance, this means that AI systems, regulatory institutions, users, and organisations all form dynamic networks of accountability, shaping how AI is perceived and accepted. ANT also highlights that governance is a continuous process of assembling and reassembling networks, meaning organisations must actively shape their AI ecosystems rather than merely conform to existing rules.

The Technology Acceptance Model (TAM) provides a framework for understanding why users adopt technology. It emphasises two primary factors: perceived usefulness (the degree to which a user believes a technology enhances their work) and perceived ease of use (the extent to which a technology is easy to operate). While TAM has traditionally focused on individual users, we extend it to organisational and transnational contexts, where regulatory compliance, ethical concerns, and social pressures shape AI adoption.

Transnational governance refers to the collective regulatory, institutional, and normative frameworks that transcend national borders (Djelic & Sahlin-Andersson, 2006). Unlike traditional governance structures that operate within national jurisdictions, transnational governance ensures that AI-powered digital innovations comply with cross-border legal, ethical, and social expectations. This governance model aligns with TAM by influencing Perceived Usefulness (through regulatory incentives and ethical guidelines) and Perceived Ease of Use (by harmonizing AI standards across different jurisdictions).

By combining these three perspectives, we argue that organisations must adopt a dual strategy, internally restructuring their AI governance while externally managing their actor-network relationships, to improve accountability and ensure ethical and market acceptance of these technologies at a transnational level.

### 3.2 Internal Reconfiguration and Optimisation of Resources

A key approach to ensuring transnational accountability is internal reconfiguration, which involves realigning organisational structures, governance mechanisms, and workforce competencies to meet evolving regulatory, ethical, and social pressures. Drawing from Latour's concept of "matters of concern" (Latour, 2004), organisations should not treat AI adoption as a mere matter of fact but rather as a socially embedded process requiring continuous accountability and ethical reflexivity. From a TAM perspective, internal reconfiguration enhances perceived usefulness by ensuring that AI systems are optimised for regulatory compliance and cultural adaptability. Embedding ethical considerations such as bias mitigation, data transparency, and fairness into AI design processes strengthens user trust and market acceptance across diverse regions (Ko and Leem, 2021; Kelly, Kaye, & Oviedo-Trespalacios, 2023). Additionally, perceived ease of use improves when governance structures facilitate seamless compliance with varying regulations, reducing friction in cross-border AI deployment (World Bank Group, 2024).

Transnational governance further requires organisations to harmonise their internal AI policies with international frameworks such as the EU AI Act, GDPR, and OECD AI principles. By proactively aligning internal practices with evolving regulatory landscapes, organisations minimise legal risks, enhance legitimacy, and shift AI adoption from a reactive response to market pressures to a deliberate, ethically informed process. To meet these demands, organisations must align governance structures, workforce skills, and technology investments with the evolving AI development and ethical standards. By fostering a culture of continuous learning and agility, organisations improve responsiveness to regulatory changes and societal expectations. Developing internal capabilities in AI ethics and compliance ensures adherence to principles such as fairness and transparency (Song, Lee, and Khanna, 2016). Moreover, optimising resource allocation leads to more efficient AI development, reducing time-to-market while meeting regulatory and market standards.

By optimising internal resources, including governance structures and workforce competencies, organisations can align AI innovations with both market and ethical expectations. This speeds up the development and deployment process, ensures regulatory compliance, and enhances trust with consumers. Organisations that integrate ethical considerations into their AI practices foster higher market acceptance, while efficient resource use strengthens competitiveness and accelerates consumer adoption (Deloitte, 2024).

In the transnational governance landscape, where AI regulations and ethical standards vary across countries, organisations must optimise internal resources to ensure compliance with global and local laws. Aligning governance structures with diverse regulatory frameworks helps organisations navigate cross-border compliance complexities, ensuring their AI-powered innovations meet ethical and legal standards across multiple jurisdictions. This approach builds global trust and enhances market acceptance in regions with differing governance structures and expectations.

### 3.3 Reshaping Boundaries and Managing Actor-Network Dynamics

The second proposed solution involves reshaping organisational boundaries and managing actor-network dynamics, acknowledging that AI-powered innovations operate within complex socio-technical networks. From the perspective of ANT, accountability is not confined to a single entity but is instead co-constructed through interactions between human and non-human actors, including regulatory bodies, consumers, AI platforms, and governance institutions. This dynamic, relational process extends beyond organisational boundaries, requiring collaborative engagement to ensure responsible AI adoption.

To enhance market acceptance and legitimacy, organisations must actively engage in collaborative AI governance with regulators, industry bodies, and civil society organisations. Participation in multi-stakeholder initiatives, such as the UN's AI for Good Summit or the Partnership on AI, ensures alignment with international ethical standards and responsiveness to evolving societal concerns. This approach builds trust, fosters compliance, and allows organisations to gain valuable insights into market needs and expectations.

Managing actor-network dynamics also helps organisations navigate transnational accountability by aligning with global AI governance structures. Rather than treating AI adoption as a top-down technological imposition, organisations must engage in public dialogues, open-source collaborations, and regulatory co-creation processes. This participatory approach enhances AI's cultural acceptance and fosters an inclusive, globally responsible AI ecosystem.

Additionally, organisations should form strategic partnerships and alliances to share resources, expertise, and influence in shaping industry standards. AI-powered innovations are not developed in isolation but within a network of social, technological, and regulatory actors. Adapting strategies to accommodate these broader socio-technical systems helps mitigate social pressures while improving market acceptance by aligning innovations with societal and stakeholder values.

From a governance perspective, organisations that proactively shape actor-network relationships can anticipate regulatory shifts, co-create ethical standards, and strengthen consumer trust. By engaging with international regulatory bodies, local governments, and global stakeholders, companies can ensure compliance with diverse cultural and regulatory environments. This collaborative approach not only fosters AI accountability but also ensures that innovations remain globally accepted, ethically sound, and aligned with transnational governance expectations.

By integrating Latourian theoretical insights with TAM and transnational governance principles, we argue that internal reconfiguration and actor-network management provide robust solutions for organisations to navigate the complex landscape of AI accountability and market acceptance. In a world where AI-driven technologies transcend national borders, organisations must move beyond passive compliance and embrace active engagement with governance ecosystems. These two solutions position organisations as responsible AI stewards, ensuring that AI-powered digital innovations are not only technologically advanced but also ethically sound, legally compliant, and socially accepted on a global scale.

## 4.0 Conclusion

The study examines the development and adoption of AI-powered digital innovations through the lens of the Technology Acceptance Model (TAM), extended to incorporate the transnational governance perspective. In particular, the study focuses on the motivations for organisations to develop and adopt AI-powered digital innovations, as well as how organisations can enhance the market acceptance of such innovations.

We recognise that AI-powered digital innovations are not inherently biased or culturally insensitive; rather, their development, deployment, and market acceptance are shaped by the organisations that control AI training platforms and direct system operations. We argue that from a transnational governance perspective, organisations must take transnational accountability for their digital innovations, ensuring they align with regulatory, ethical, and cultural expectations. Regulatory and ethical pressures serve as key factors driving technology acceptance, compelling organisations to comply with social norms and legal frameworks. However, ensuring ethical AI use is not the responsibility of any single entity; instead, multinational organisations, international regulatory bodies, and cross-border advocacy groups collectively shape AI governance. Social pressures, including AI regulations, ethical considerations, and cultural acceptance, play a crucial role in motivating organisations to advance technology while fostering market acceptance and mutual accountability. To navigate these challenges, organisations must adopt a dual strategy: internally restructuring governance through resource reconfiguration while externally managing actor-network relationships. By integrating Latourian theoretical insights, TAM, and transnational governance principles, we propose that internal reconfiguration and actor-network management offer robust solutions to strengthen AI accountability and acceptance. In an era where AI operates beyond national borders, organisations must move beyond passive compliance and engage proactively in governance ecosystems to ensure AI is technologically innovative, ethically sound, legally compliant, and socially accepted on a global scale.

Future research could expand upon this work in several promising directions. First, empirical dataset could be developed to support the theoretical framework proposed in this study. Second, this work proposed internal resource reconfiguration; however, we must highlight that the infrastructure accretes during this process (Power 2015). Future research could investigate the alignment between organisational strategies and infrastructure accretion development, in particular information and financial infrastructure (Tan, Abdaless and Liu 2018; Tan, Liu and White 2013). Third, this work also proposed reshaping boundaries and managing the actor-network dynamics as a solution. Future research might examine the mechanisms and search strategies to manage such actor-network dynamics. Fourth, there is a difference between market acceptance and consumer affordance (El Amri ans Akrout 2020; Hafezieh and Eshraghian 2017). Future research can discuss the opportunities and challenges of turning AI-powered digital innovations into affordable products for consumers. Fifth, whilst this study outlines several strategies conceptually, translating them into concrete steps requires further investigation. Further research should focus on developing practical, actionable guidelines to support organisations in implementing the proposed solutions of resource reconfiguration and actor-network management. Empirical case studies might examine best practices for restructuring internal governance mechanisms to enhance AI accountability, as well as frameworks for effectively managing actor-network relationships across transnational regulatory environments.

Finally, it is important to acknowledge that AI-powered digital innovations vary significantly in their design and functionality. For example traditional symbolic or rules-based AI systems operate on a predefined logic, making them relatively predictable but limited in their adaptability, whereas modern neural AI models – such as those used in generative AI and predictive learning systems – leverage vast datasets and continuous learning, often rendering them as black box systems, impenetrable to human understanding (Zednik, 2021). These differences will also likely influence how organisations develop and adopt these technologies, which provides a further promising avenue for future research.